\documentclass[journal=jacsat,manuscript=article]{achemso}

\usepackage[version=3]{mhchem} 
\usepackage{booktabs} 
\usepackage{multirow}
\usepackage{hyperref}
\hypersetup{colorlinks=true, linkcolor=blue, citecolor=blue, urlcolor=blue} 

\newcommand*\chem[1]{\ensuremath{\mathrm{#1}}} 

\author{Raagya Arora}
\email{raagya@g.harvard.edu}
\affiliation[Harvard SEAS]
{John A. Paulson School of Engineering
and Applied Sciences, Harvard
University, Cambridge, Massachusetts 02138, United States}
\altaffiliation{These authors contributed equally to this work.}
\author{Ariel R. Barr}
\altaffiliation{These authors contributed equally to this work.}
\affiliation[Massachusetts Institute of Technology]
{Department of Materials Science \& Engineering, Massachusetts Institute of Technology, Cambridge, Massachusetts 02139, United States}
\author{Daniel Bennett}
\affiliation[Harvard SEAS]
{John A. Paulson School of Engineering
and Applied Sciences, Harvard
University, Cambridge, Massachusetts 02138, United States}
\author{Daniel T. Larson}
\affiliation[Harvard University]
{Department of Physics, Harvard
University, \\ Cambridge, Massachusetts 02138, United States}
\author{Michele Pizzochero}
\affiliation[Harvard SEAS]
{John A. Paulson School of Engineering and Applied Sciences, Harvard
University, Cambridge, Massachusetts 02138, United States}
\author{Efthimios Kaxiras}
\affiliation[Harvard University]
{Department of Physics, Harvard
University, \\ Cambridge, Massachusetts 02138, United States}
\alsoaffiliation{John A. Paulson School of Engineering
and Applied Sciences, Harvard
University, Cambridge, Massachusetts 02138, United States}
\email{kaxiras@g.harvard.edu}

\title[An \textsf{achemso} demo]
  {
  2D Nitride Ordered Alloys: \\ A Novel Class of Ultra-Wide Bandgap Semiconductors
  }

\abbreviations{IR,NMR,UV, UWBG, CBM, VBM, 2D}

\begin{document}









\newpage
\begin{abstract}
  Ultra-wide bandgap (UWBG) semiconductors are poised to transform power electronics by surpassing the capabilities of established wide bandgap materials, such as GaN and SiC, owing to their capability to operate at higher voltage, frequency, and temperature ranges. While bulk group-III nitrides and their alloys have been extensively studied in the UWBG realm, their two-dimensional counterparts remain unexplored. Here, we examine the stability and electronic properties of monolayers of ordered boron-based group-III nitride alloys with general formula B$_{\rm x}$\emph{M}$_{1-\rm x}$N, where \emph{M} = Al, Ga. On the basis of \emph{ab initio} calculations we identify a number of energetically and dynamically stable structures. Instrumental to their stability is a previously overlooked out-of-plane displacement (puckering) of atoms, which induces a polar ordering and antiferroelectric ground state. 
  Our findings reveal the energy barrier between metastable ferroelectric states is lowered by successive switching of out-of-plane displacements through an antiferroelectric state.
\end{abstract}
\clearpage 
Ultra-wide bandgap (UWBG) semiconductors have recently emerged as a promising class of materials with the potential to advance power electronics, optoelectronics, and radio frequency devices\cite{tsao2018ultrawide}.
%
%
Defined by a bandgap $\epsilon_{\rm{gap}}$ exceeding that of gallium nitride (GaN, $\epsilon_{\rm{gap}}=$ 3.4 eV), UWBG materials are resistant to electrical breakdown and exhibit high thermal conductivity and mechanical strength, which enable them to handle high voltages and efficiently dissipate heat. 
For decades, the landscape of UWBG semiconductors employed in power- and opto-electronics has been dominated by established bulk materials. 
Examples include \chem{Ga_2O_3}, diamond, MgZnO, indium tin oxide, indium gallium zinc oxide, and group-III nitrides (GaN, AlN, \chem{Al_xGa_{1-x}N})\cite{tsao2018ultrawide, yang20222d, wong2021ultrawide}. 

As Moore's law nears its end, traditional UWBG semiconductors will face challenge to meet the demands of future high-performance (opto)electronic devices. 
In response, research is shifting towards low-dimensional materials with novel structures and adaptability. 
Fueled by the discovery of graphene in 2004\cite{novoselov2004electric}, the field of 2D materials has experienced exponential growth, yielding promising candidates for 2D UWBG semiconductors that include group-III metal chalcogenides (GaS, GaSe), metal oxyhalides, metal nitrides, and metal oxides\cite{wong2021ultrawide}. 
These materials offer unique possibilities for design and fabrication of next generation high-power transparent electronic devices, deep-ultra-violet photodetectors, flexible electronic skins, and energy-efficient displays\cite{yang20222d,noor2023engineering}.

In the present work we focus on boron-based group-III nitrides in 2D structures, since bulk hexagonal boron nitride (hBN) is one of the most stable known UWBG materials. hBN exhibits high thermal conductivity, high chemical stability, mechanical strength, and high quality single crystal and multilayer growth\cite{falin2017mechanical, watanabe2004direct}. Other variants, including hexagonal aluminium nitride (hAlN) and gallium nitride (hGaN), also exist in a planar structure and exhibit exceptional electronic, optical, and thermoelectric properties\cite{ben20212d}. While hBN has been used extensively as an encapsulating layer and gate dielectric for 2D electronics\cite{dean2010boron,xue2011scanning}, high quality \emph{n}-type or \emph{p}-type doping of hBN has not been reproducibly achieved\cite{lyons2014effects}. 
The fundamental difficulty in doping hBN originates from the high conduction band minimum (CBM) and low valence band maximum (VBM) relative to the vacuum energy level, as well as heavy hole effective masses, making it susceptible to the formation of small polarons. 
Consequently, impurities in hBN favor the formation of deep, localized, mid-bandgap levels, resulting in defect states robust against efficient ionization. 
As a result, both \emph{p}- and \emph{n}-type doping is extremely difficult to achieve in hBN via conventional methods\cite{majety2013electrical,weston2018native}. 
Here, in order to address this issue, we consider alloys of hBN where B is partially replaced by Al or Ga.

The combination of bulk AlN and GaN yields the UWBG alloys \chem{Ga_xAl_{1-x}N}\cite{nam2004unique}. Bulk
AlN exhibits a significantly wider bandgap (6 eV) than GaN (3.4 eV), so the combination results in tunable bandgaps across a broad energy range for \chem{Ga_xAl_{1-x}N} alloys. 
These bulk alloys offer additional advantages including exceptional breakdown fields exceeding 10 MV cm$^{-1}$, and high electron mobility reaching up to 1000 cm$^{2}$ V$^{-1}$ s$^{-1}$. 
Further, \emph{n}-type doping with Si is easily accomplished in bulk \chem{Ga_xAl_{1-x}N} due to its low donor ionization energy\cite{jiang2002algan}. 

In the case of B-based group III nitrides, the structural and bonding environment presents significant challenges arising from the inherent differences in the preferred bulk crystal structures of BN and AlN or GaN:\cite{PhysRevMaterials.1.065001} 
BN exhibits a thermodynamically stable hexagonal phase, whereas AlN and GaN adopt a wurtzite structure. 
This disparity hinders the straightforward incorporation of boron into the wurtzite structures of AlN or GaN. 
Recent efforts have focused on overcoming this challenge by developing thin films of wurtzite \chem{B_xAl_{1-x}N} on substrates like AlN and sapphire\cite{calderon2023atomic}. 
Earlier computational studies \cite{milne2023ab, milne2023electronic, zhang2017structural} offer a deeper understanding of the structure and bonding environment of bulk wurtzite \chem{B_xAl_{1-x}N}. In this work we explore the stable, two-dimensional alloy phases of hBN, namely  \chem{B_x\emph{M}_{1-x}N}  with \emph{M}=Al or Ga.
%
 
While other 2D hexagonal group-III nitride alloys such as \chem{Ga_xAl_{1-x}N} and \chem{Ga_xIn_{1-x}N} have been theoretically proposed, stable 2D B-based nitride alloys have not been demonstrated\cite{chen2023two}. 
Previous works focused on planar geometries and did not thoroughly explore possible thermodynamically and dynamically stable crystal structures. 
Here we identify a critical factor for the stability of 2D B-nitride alloys, namely out-of-plane distortion (puckering) of the crystal structure, which can impact stability, enhance spin-orbit coupling, and break the sublattice symmetry associated with the isotropic planar structure, thus modifying the electronic, thermal, and topological properties\cite{reis2017bismuthene, quhe2012tunable}. 
For example, puckered monolayer 2D Xenes such as silicene exhibit exotic electronic and magnetic properties \cite{tao2015silicene} such as giant magnetoresistance\cite{xu2012giant}, chiral superconductivity\cite{liu2013d+}, and the quantum spin Hall effect \cite{liu2011quantum}. 
The highly corrugated structure of 2D black phosphorus similarly results in a novel response to applied stimuli such as strain, electric field, or polarized light, which do not occur in the bulk structure\cite{xia2014rediscovering, he2015exceptional, phaneuf2016polarization}. In a similar vein we find that the puckered structures of \chem{B_x\emph{M}_{1-x}N} exhibit an unusual antiferroelectric (AFE) ground state, through which polarization switching between metstable ferroelectric (FE) states occurs.
This phenomenon gives rise to a three-state system with potential applications in low-dimensional non-volatile memory devices\cite{qi2021review}. The competition between FE and AFE ordering within the B-based alloys allows the superstructure to persist even in monolayer crystals. 


To gain a detailed understanding of the properties of B-based nitride alloys and motivate their experimental realization, we present an {\it ab initio} study of their structural and electronic properties using semi-local and hybrid density functional theory (DFT) calculations. 
 We consider the B-based group-III nitrides B$_{\rm x}$\emph{M}$_{1-\rm x}$N where \emph{M} = Al, Ga,  for x = 0, 0.25, 0.50, 0.75, 1, using 2D hexagonal structures in  $2\times 2$ and $4\times 4$ supercells of the primitive cell of hBN. 
The thermodynamic stability of an alloy is determined by the formation energy, defined by $E_{\rm{f}} = E_{\text {tot }}-\sum_{\mathrm{n}} \mu_{\mathrm{n}} \eta_{\mathrm{n}}$ where $E_{\text {tot }}$ is the total energy of the 2D alloy structure, $\eta_{\mathrm{n}}$ is the concentration of element n in the alloy, and $\mu_{\mathrm{n}}$ is the chemical potential of element n, which by convention is taken to be the total energy per atom of the constituent elements (Al, B, N, Ga) in their most stable reference state. The thermodynamic stability of the various alloys we consider is summarized in Table 1. 

\begin{table}[h!]
\captionsetup{font=footnotesize}
\centering
\caption{Evolution of the formation energy (E$_{\rm{f}}$) and bandgap ($\epsilon_{\rm{gap}}$) of 2D nitride ordered alloys, in either planar or puckered geometry, as a function of an increasing boron content (x)}
\centering
\resizebox{\textwidth}{!}{
\label{table:qho_table}
\begin{tabular}{l l l c c c c c c c c c c c c }
\toprule
Material &  & x = \multirow{2}{*}{ }
 & 0.00 & & \multicolumn{2}{c}{0.25} &  & \multicolumn{2}{c}{0.50} & & 0.75 &  & 1.00       \\  \cmidrule{4-4} \cmidrule{6-7}  \cmidrule{9-10} \cmidrule{12-12} \cmidrule{14-14}
   & &  & planar &  & planar & puckered & & planar & puckered & & planar  & &  planar     \\ 
\midrule

B$_{\rm x}$Al$_{1-\rm x}$N  &  E$_{\rm{f}}$ & & $-$0.97 &   & $-$1.25 & $-$1.26 &   & $-$1.10 & $-$1.11  & & $-$1.07 &  & $-$1.25 &  \\

 & $\epsilon_{\rm{gap}}$ & & 4.04 &   & 4.09 &  4.20 &  & 4.52 & 4.73  & & 5.16 &  & 5.71 & \\
\cmidrule{1-2}
B$_{\rm x}$Ga$_{1-\rm x}$N  & E$_{\rm{f}}$  & & $-$0.16 &   & $-$0.23 & $-$0.24  &   & $-$0.27 & $-$0.30  & & $-$0.63 &  & & \\

 &  $\epsilon_{\rm{gap}}$ & & 3.76 &   & 3.76 & 4.45 &   & 3.96 & 4.17  & & 5.26& & & \\


\bottomrule
\end{tabular}} 

\end{table}

All hexagonal B$_{\rm x}$Al$_{1- \rm x}$N and B$_{\rm x}$Ga$_{1-\rm x}$N alloys we consider have negative formation energies, which is a necessary condition for stability. Among the pure binary hexagonal Al/B/Ga nitrides (point group D$_{3h}$, space group \#194, P6$_3$/mmc), hBN exhibits the lowest $ E_{\rm{f}}$, followed by AlN and then GaN. This trend aligns with the known stability of hexagonal phases of these materials, in comparison to the  wurzite structures. For a fixed composition x, B$_{\rm x}$Al$_{1- \rm x}$N alloys consistently display lower $E_{\rm{f}}$ values compared to B$_{\rm x}$Ga$_{1- \rm x}$N alloys.  

 
\begin{figure}
        \centering
        \includegraphics[width=1\linewidth]{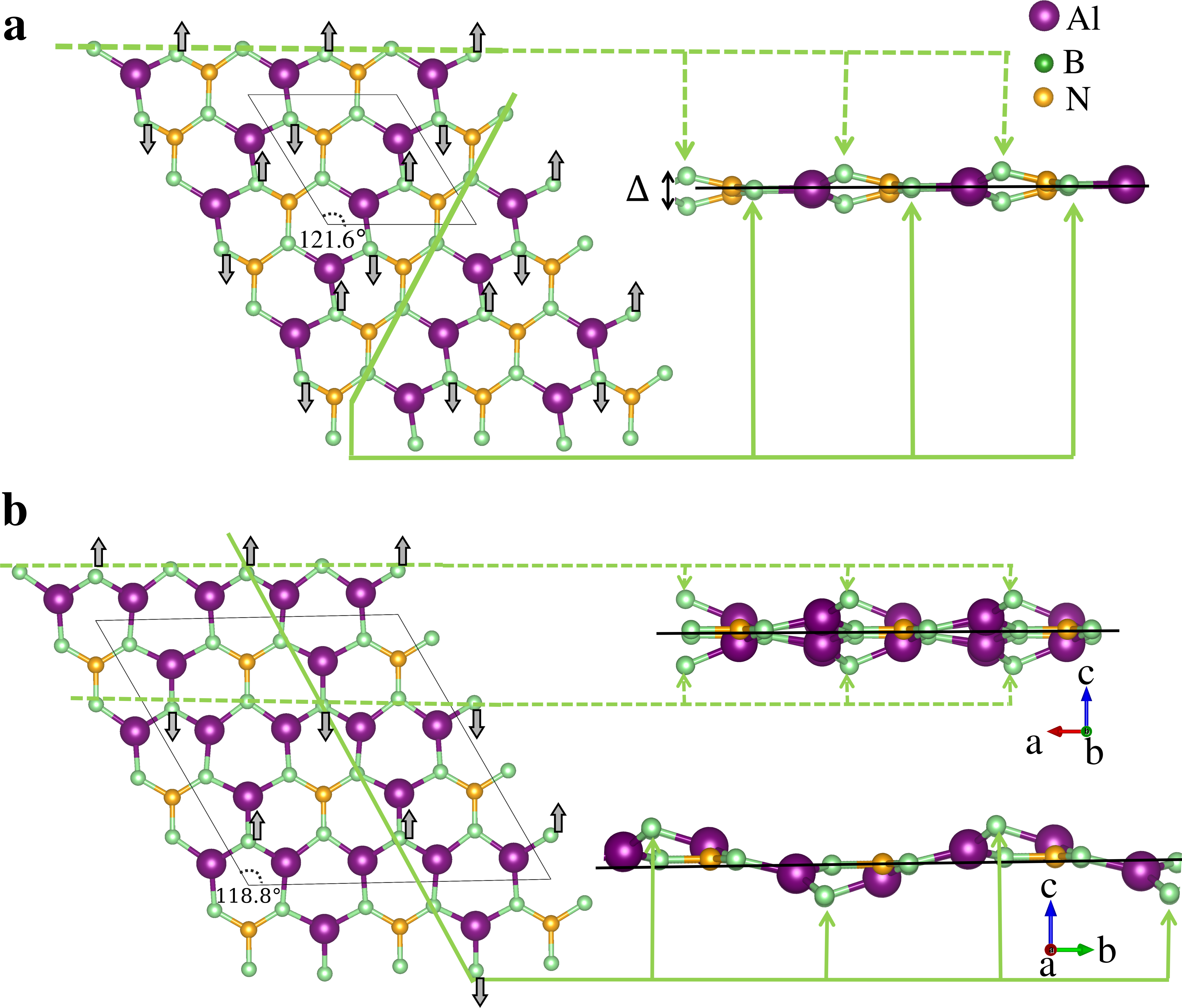}
        \caption{Top and side views of stable 2D polar hexagonal monolayers of B$_{\rm x}$Al$_{1-\rm x}$N with Al concentration   {\bf (a)} x = 0.5,  {\bf (b)} x = 0.75. The up and down arrows indicate the out-of-plane displacements ($\Delta$) of N atoms relative to the ideal planar geometry. The 2$\times$2 supercell for x = 0.5 and the 4$\times$4 supercell for x = 0.75 are highlighted in black. 
        }
        \label{fig:Figure1} 
    \end{figure}
    
To assess the dynamical stability of group-III nitride alloys in 2D form, we calculate the phonon spectra using density functional perturbation theory (DFPT). For the pure phases BN, AlN, and GaN, the calculated phonon spectra show no imaginary frequencies (see Fig.~S1 in Supplementary Information, SI), confirming their dynamical stability in the hexagonal planar configuration. Furthermore, these pristine III-V nitride monolayers exhibit insulating behavior in their planar geometries, with indirect band gaps of 5.71 eV, 4.05 eV, and 3.76 eV for BN,  AlN, and GaN, respectively, as calculated with the HSE hybrid functional. The prevalence of strong $sp^2$ bonding in these structures allows the group-III cations to adopt a planar geometry.

\begin{figure}
    \centering
    \includegraphics[width=0.8\linewidth]{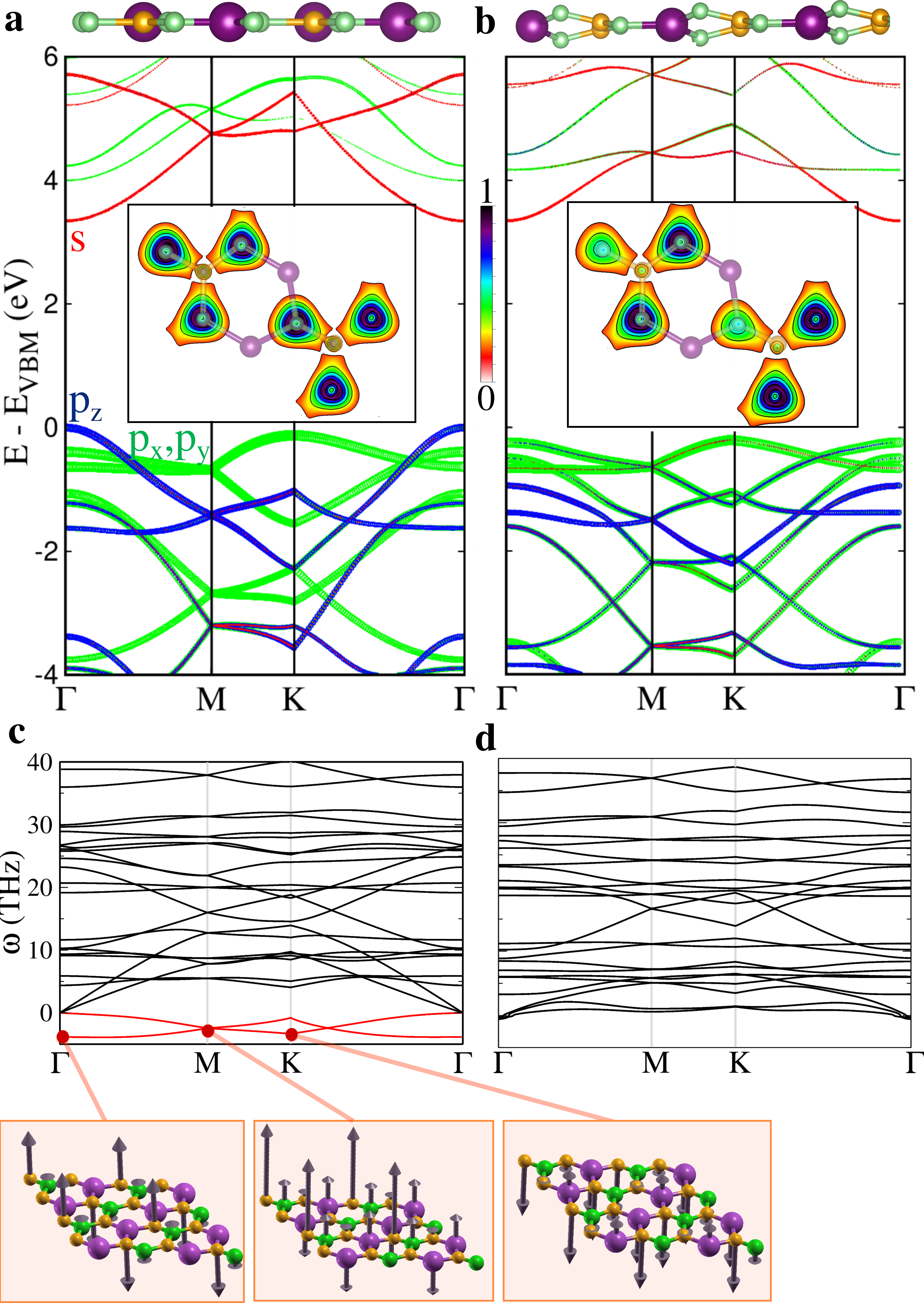}
    \caption{Electronic structure of \chem{B_{0.5}Al_{0.5}N} projected onto $s$ and $p$ orbitals of N-atoms:  {\bf (a)} hexagonal planar and  {\bf (b)} distorted hexagonal puckered structures. {\bf (c)} Phonon dispersion of the planar structure reveals significant instability throughout the Brillouin zone (red lines) absent in the stable puckered structure {\bf (d)}. The three panels below {\bf (c)} show the atomic displacements due to unstable phonon modes at various high-symmetry points indicated by red dots. The insets in (a) and (b) show 2D charge density maps for the planar and puckered \chem{B_{0.5}Al_{0.5}N}. }
    \label{fig:Figure2}
\end{figure}

On the other hand, 2D alloys of BN with Al or Ga adopt a puckered structure. Fig.~\ref{fig:Figure1} shows two specific examples: \chem{B_{0.5}Al_{0.5}N} and \chem{B_{0.25}Al_{0.75}N}.
The hexagonal planar structure of B$_{0.5}$Al$_{0.5}$N has a wide band gap of 4.7 eV. The atom-projected electronic structure is presented in Fig.~\ref{fig:Figure2}(a) and Fig.~S2, and exhibits well separated valence bands composed of Al and B atom $s$ and $p$ orbitals, while the conduction band is composed of N $s$ and $p$ orbitals. The valence band at $\Gamma$ primarily consists of N atom out-of-plane $p_z$ orbitals while the valence band at the M and K points is composed of in-plane N $p_x$,$p_y$ orbitals. The interaction between $p_z$ orbitals and $p_x$,$p_y$ orbitals of anions in the planar lattices vanishes due to the horizontal reflection symmetry.
The inset of Fig.~\ref{fig:Figure2}(a) shows the calculated charge density plot, which confirms the prevalence of $sp^2$ bonding in these structures that allows the B cations to adopt a trigonal planar geometry. Achieving and maintaining a planar geometry in B-based alloys at different concentrations relies on strong $\pi$-bonding between the $p_z$ obitals\cite{csahin2009monolayer}, similar to graphene and hBN.

Fig.~\ref{fig:Figure2}(c) shows the phonon spectrum of planar B$_{0.5}$Al$_{0.5}$N.
The presence of imaginary frequencies throughout the spectrum (red lines) indicates dynamical instability in this structure. Further analysis of the phonon eigenvectors at $\Gamma$ confirms that the soft modes correspond to the out-of-plane displacements of the N atoms with either the same or opposite signs in the zig-zag and arm chair directions, respectively.  These soft modes act as a driving force, spontaneously transforming the planar B$_{0.5}$Al$_{0.5}$N structure into a lower-energy stable phase.  The dynamical stability of these new puckered phases is confirmed by their phonon spectra, shown in Fig.~\ref{fig:Figure2}(d) and Fig.~S3, which display no imaginary frequencies. Polar ordering emerges as a consequence of the N atom distortions along the out-of-plane direction, which stabilize an anti-ferroelectric (AFE) phase of B$_{0.5}$Al$_{0.5}$N. Consistent with the soft modes observed at $\Gamma$ in the planar structure, in the puckered structure the N atoms bonded to two Al atoms are displaced in opposite directions, while the remaining two N atoms which are bonded to two B atoms prefer to remain in-plane. The stable puckered structure of B$_{0.5}$Al$_{0.5}$N is an hexagonal structure with lower symmetry, with lattice constants  $a$=5.67, $b$= 5.53 \AA{} and angle $\gamma = 121.6^{\circ}$ that deviates slightly from 120$^{\circ}$ (Fig.~\ref{fig:Figure1}(a)).  We carried out a similar analysis of the structures of nitride alloys with gallium instead of Al (see SI for details).

\begin{figure}
    \centering
    \includegraphics[width=0.85\linewidth]{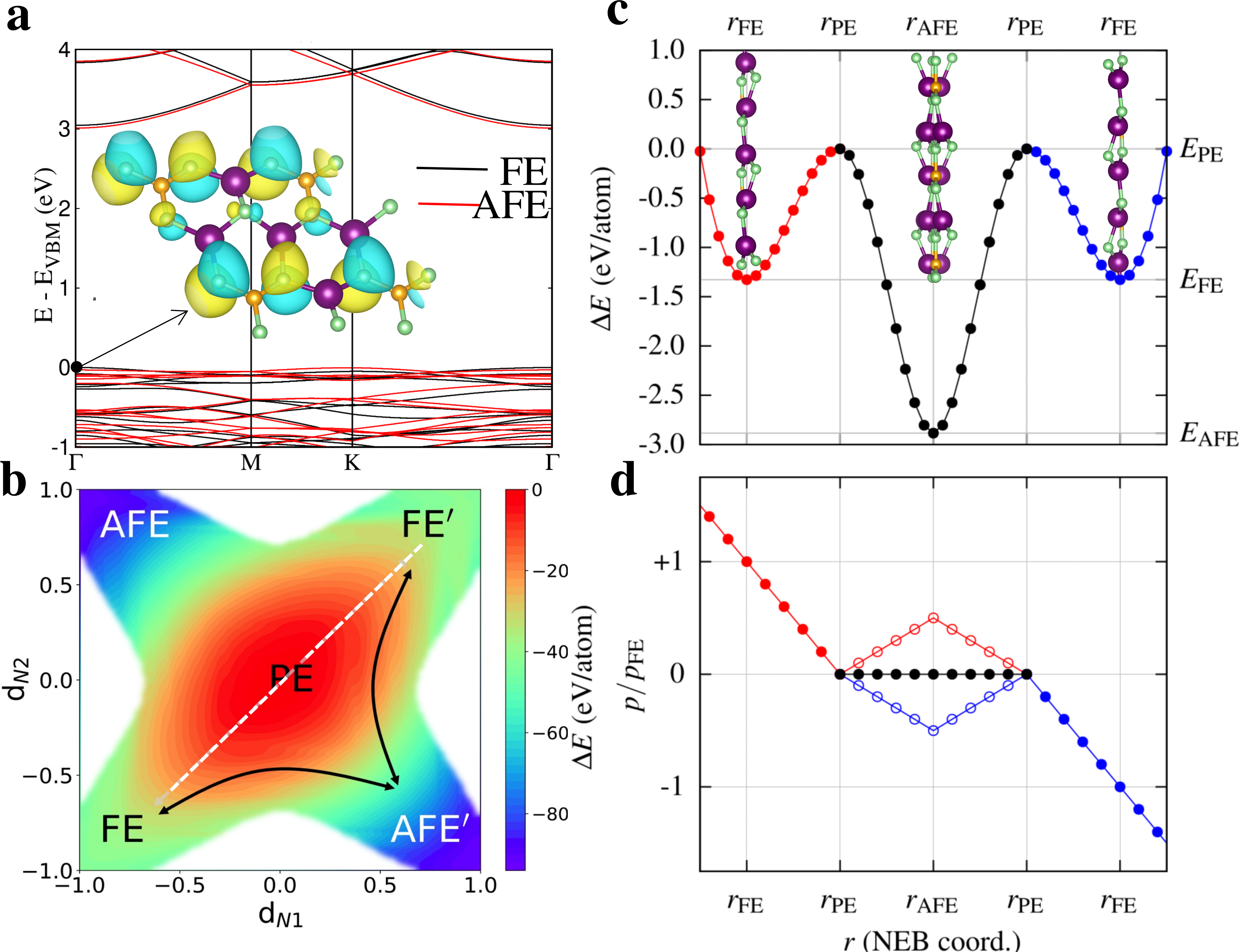}
    \caption{
    {\bf (a)} Electronic band structure of \chem{B_{0.25}Al_{0.75}N} in the FE (black) and AFE (red) phases. The inset shows charge density isosurfaces of the VBM state at $\Gamma$ for the AFE phase.
    {\bf (b)} Difference in energy with respect to the PE phase as a function of the out-of-plane displacement of the N atoms in a $2\times 1$ supercell. The displacements $d_{{\rm N}_i}$, $i=1,2$, are in units of the equilibrium displacement for the FE/AFE phases. The dashed white line indicates the energetically costly switching path through the PE phase, while the solid black arrows denote a lower-energy switching path through an AFE phase.
    {\bf (c)} Total energy per atom as a function of NEB coordinate $r$ (b). The FE and AFE structures are shown in insets. The FE and AFE phases are separated by the dynamically unstable PE phase. The red and blue points indicate polarization with positive and negative signs, respectively, and the black points indicate zero total polarization.
    {\bf (d)} Out-of-plane dipole moment as a function of NEB coordinate $r$, in units of the dipole moment of the FE phase. The filled symbols indicate the total dipole per unit cell, and the open symbols in the AFE region show the local contributions in each cell, which are equal and opposite, verifying that the phase is indeed AFE.
    }
    \label{fig:Figure3}
\end{figure}

Puckering of the lattice lowers the energy by 7 meV/atom and further opens the bandgap (Fig.~\ref{fig:Figure2}(b) and Fig.~S6 of SI). We present a detailed analysis of the orbital-projected electronic structure of B$_{\rm x}$Al$_{1-\rm x}$N, (see Fig.~\ref{fig:Figure2} and Fig.~S6) at various concentrations. As is clear in Fig.~\ref{fig:Figure2}(b), the out-of-plane distortion leads to notable changes in the bands with N $p_z$ and $p_x$, $p_y$ character. Puckering lowers the lattice symmetry and permits previously symmetry-forbidden mixing between $p_z$ and $p_x$, $p_y$ orbitals, which pushes the N-$p_z$-derived orbitals to lower energy and increases the bandgap. The origin of puckering at high concentrations of Al/Ga (x$<$0.75) in hBN alloys can also be understood in terms of the atomic radii of B, Al, and Ga, which are 85, 125, and 130 pm, respectively. Large atomic size differences between Al/Ga and B atoms in their trigonal environment introduce a heterogeneous bonding environment, with the ensuing strain being relieved by the lower energy puckered configuration.

Puckering also affects the character of the chemical bonding \cite{csahin2009monolayer}. 
Strong $\pi$-bonding between the $p_z$ orbitals of B and N atoms is crucial for maintaining a stable planar geometry, but as the concentration of the larger Al atoms increases ($x<0.75$), the average bond length between nearest neighbors increases. For example, at $x = 0.5$, the average B-N bond length changes from 1.47 \AA{} in the planar geometry to 1.51 \AA{} in the puckered lattice. This increase in bond length weakens the $\pi$-bonding due to the decreased overlap of $p_z$ orbitals, and destabilizes the planar geometry in B$_{\rm x}$Al$_{1-\rm x}$N alloys. The inset of Fig.~\ref{fig:Figure3}(a) shows the isosurfaces of charge density associated with the VBM at $\Gamma$ for the puckered alloy with $ x=0.25$, demonstrating the weaker overlap between neighbouring $p_z$ orbitals. Thus, puckering disrupts the planar $sp^2$ hybridization, causing the $s$, $p_x$, and $p_y$ orbitals to recombine with the $p_z$ orbital to form orbitals closer to $sp^3$-like hybridization. The weak interaction between the hybrid orbitals is reflected in the dispersion of the corresponding electronic bands, explaining the occurrence of flat valence bands in the puckered electronic structure of \chem{B_{0.25}Al_{0.75}N} and \chem{B_{0.25}Ga_{0.75}N} (see Fig.~2b).

The periodic supercell used for structural optimization must be large enough to capture the atomic relaxation in the ground state structure of B$_{0.25}$Al$_{0.75}$N.
While a 2$\times$2 supercell hosts a metastable ferroelectric (FE) state (Fig.~S7), a larger 4$\times$4 supercell reveals a lower-energy AFE ground state (Fig.~\ref{fig:Figure1}(b)), overlooked in earlier studies\cite{noor2023engineering}.
The interplay between out-of-plane atomic displacements leads to interesting polar ordering in these alloys, namely a metastable FE state and a stable AFE ground state in \chem{B_{0.25}Al_{0.75}N}. The electronic band structures for both phases are similar, as shown in Fig.~\ref{fig:Figure3}(a). However, the AFE phase, where N atoms in neighboring cells have an alternating out-of-plane displacements, is approximately a factor of two lower in energy than the FE phase, with respect to the planar paraelectric (PE) phase, see Fig.~\ref{fig:Figure3}(b) and (c). In the FE phase the inversion symmetry is broken by the out-of-plane displacements of the N atoms, which results in a dipole moment $\vec{p}_{z}$. The dipole moment as a function of out-of-plane displacement is obtained by calculating the Berry phases\cite{vanderbilt1993electric,king1993theory} from DFT, see Fig.~\ref{fig:Figure3}(d). The computed energy versus buckling displacement $u$ (see Fig.~S7) of \chem{B_{0.25}Al_{0.75}N} reveals a double-well potential, with the dipole moment given in terms of the equilibrium value in the FE phase, $p_{\rm FE} = $ 4.83$\times10^{-22} \; \rm C \cdot \rm m$. The AFE phase is characterized by an alternating pattern of out-of-plane displacements and has considerably lower energy than the FE phase. A zero total dipole moment is found for every alternating pattern of displacements. In order to distinguish the AFE phase from a PE phase, we calculate the local polarization \cite{bennett2023theory} in each cell using the Born effective charges \cite{ghosez1998dynamical,bennett2024asymmetric}, defined as 
\begin{equation}
Z^{*}_{\kappa,\alpha\beta} = \frac{\partial{p_{\alpha}}}{\partial u_{\kappa,\beta}}
\end{equation}
where $u_{\kappa,\beta}$ is the displacement of atom $\kappa$ in direction $\beta$ with respect to the nonpolar configuration. $Z^{*}$ is approximately constant as a function of displacement, and the local dipole (Fig.~\ref{fig:Figure3} (d)) in each cell $i$ can therefore be calculated as 
\begin{equation} 
p^{i}_z = \sum_{\kappa\in i,\beta} Z^{*}_{\kappa,z\beta} u_{\kappa,\beta}, \end{equation} 
where the sum over atoms is confined to cell $i$.
In the AFE phase there are nonzero local dipole moments in each cell, which are equal and opposite and half the magnitude of the dipole moment in the FE phase (Fig.~\ref{fig:Figure3} (d)).

In FE materials, the direction of the spontaneous polarization can be switched using an external electric field, hence it is important to understand the kinetics of the polarization reversal process. In order to explore the most favorable transition pathway connecting the two energetically degenerate FE states with opposite polarization directions, we systematically scanned the energy surface with respect to two independent reaction coordinates, $d_{{\rm N}_1}$ and $d_{{\rm N}_2}$, which describe the polar displacements of distinct N atoms in the individual cells.
Rather than switching polarization through the PE state, the large energy barrier is avoided by passing through the AFE state, as indicated by the black arrows in Fig.~\ref{fig:Figure3}(b). We used the nudged elastic band (NEB) method to identify the transition pathways and determine the energy barriers between the different ferroelectric states. 
The energy barrier for switching is suﬃciently high to ensure the presence of three distinct states that could be exploited for nonvolatile memory applications: $(+p_z, 0, -p_z)$. 

Similar to our findings, the competition between AFE and FE order has been predicted in other 2D materials such as CuInP$_2$S$_6$\cite{chandrasekaran2017ferroelectricity}. 
Although a large energy barrier is predicted by DFT, in realistic scenarios the barrier is typically orders of magnitude lower because switching occurs through the nucleation and motion of domain walls. The description of domain walls in these alloys is beyond the scope of this work and is left as a subject of future research.

In conclusion, our first-principles investigation into ordered two-dimensional hexagonal group-III nitride alloys (B$_{\rm x}$\emph{M}$_{1-\rm x}$N with \emph{M} = Al, Ga), predicts out-of-plane distortions in the stable structures of these materials. In addition to greater stability and larger band gaps, the puckered structure also leads to interesting polar ordering.
Thus 2D boron-based hexagonal nitride alloys present a promising class of materials for the design of high-performance UWBG electronics with superior operating voltages, frequencies, and efficiencies, as well as intriguing electromechanical properties.

%
%
%

%

\begin{acknowledgement}

The authors thank Tomas Palacios,  Jing Kong, Jagadeesh Moodera and their research groups for valuable discussions.
This work was supported mainly by Army Research Office Agreement Number W911NF-23-2-0057,
and partly by ARO Agreement Number W911NF-21-1-0184 and ARO Cooperative Agreement Number W911NF-21-2-0147.
Calculations were performed on the FASRC cluster supported by the FAS Division of Science Research Computing Group at Harvard University.

\end{acknowledgement}

\begin{suppinfo}
Technical details and parameters of first-principles calculations along with additional results of the structural and electronic properties of 2D nitride ordered alloys and related systems.

\end{suppinfo}

\providecommand{\latin}[1]{#1}
\makeatletter
\providecommand{\doi}
  {\begingroup\let\do\@makeother\dospecials
  \catcode`\{=1 \catcode`\}=2 \doi@aux}
\providecommand{\doi@aux}[1]{\endgroup\texttt{#1}}
\makeatother
\providecommand*\mcitethebibliography{\thebibliography}
\csname @ifundefined\endcsname{endmcitethebibliography}  {\let\endmcitethebibliography\endthebibliography}{}

\end{document}